\documentclass[aps,prc,twocolumn,noshowpacs, superscriptaddress,floatfix]{revtex4}

\usepackage{amsmath}
\usepackage{amssymb}
\usepackage{CJK}
\usepackage{graphicx}
\usepackage{dcolumn}
\usepackage{bm}
\usepackage{color}
\usepackage{ulem}

\begin{document}
%
\begin{CJK*}{GB}{gbsn}
\title{Finite-size scaling phenomenon of nuclear liquid--gas phase transition probes}

\author{H. L. Liu}
\affiliation{Shanghai Institute of Applied Physics, Chinese Academy of Sciences, Shanghai 201800, China}
\affiliation{University of the Chinese Academy of Sciences, Beijing 100080, China}
\affiliation{School of Physical Science and Technology, ShanghaiTech University, Shanghai 201203, China}

\author{Y. G. Ma}\thanks{Corresponding author. Email: mayugang@cashq.ac.cn}
\affiliation{Key Laboratory of Nuclear Physics and Ion-beam Application (MOE), Institute of Modern Physics, Fudan University, Shanghai 200433, China}
\affiliation{Shanghai Institute of Applied Physics, Chinese Academy of Sciences, Shanghai 201800, China}

\author{D. Q. Fang}
\affiliation{Key Laboratory of Nuclear Physics and Ion-beam Application (MOE), Institute of Modern Physics, Fudan University, Shanghai 200433, China}
\affiliation{Shanghai Institute of Applied Physics, Chinese Academy of Sciences, Shanghai 201800, China}

\date{\today}

\begin{abstract}
 {Based on the isospin-dependent quantum molecular dynamics model, finite-size scaling effects on nuclear liquid--gas phase transition probes are investigated by studying the de-excitation processes of six thermal sources of different sizes with the same initial density and similar $N/Z$. Using several probes including the total multiplicity derivative ($dM_{tot}/dT$), second moment parameter ($M_2$),  intermediate mass fragment (IMF) multiplicity ($N_{IMF}$), Fisher's power-law exponent ($\tau$), and Ma's nuclear Zipf's law exponent ($\xi$), the relationship between the phase transition temperature and the source size has been established. It is observed that the phase transition temperatures obtained from the IMF multiplicity, Fisher's exponent, and Ma's nuclear Zipf's law exponent have a strong correlation with the source size. Moreover, by employing the finite-size scaling law, the critical temperature $T_c$ and the critical exponent $\nu$ have been obtained for infinite nuclear matter.}
\end{abstract}


\maketitle

\section{INTRODUCTION}
\label{introduction}
Exploration of the properties of nuclear matter is a hot topic in nuclear physics research. It is associated with the nuclear equation of state (EOS) and the dynamical description of the heavy-ion collision process ~\cite{gghz2014-1,blc2008-2,bmf2008-3,Ma2,ygm2004-5}. The phenomenon of liquid--gas phase transition (LGPT) is observed in water; when the temperature reaches a certain value, water changes from liquid phase to gas phase. 
The similarity between the shape of the nucleon-nucleon interaction in
nuclear matter and the van der Waals nature of the intermolecular
interaction in water indicates the possibility of occurrence of LGPT
in nuclear matter, similar to that in water.
This phase transition feature has attracted significant attention from the heavy-ion collision community at relativistic ~\cite{ANSS2005-7,ZXCG2008-8,Chen,PBM,Luo,Song,Huang} and intermediate energies ~\cite{Veselsky,samb2001-11,jpo1995-12,ygm1997-14,ygm1999-15}. In particular, the study of LGPT in intermedium energy heavy-ion collisions is a current focus in both experimental and theoretical studies ~\cite{gghz2014-1,blc2008-2,bmf2008-3,Ma2,ygm2004-5,samb2001-11,jpo1995-12,ygm1997-14,ygm1999-15,JLC2013-13,ygm2002-16,DXG2016-17,JBLG2002-33,ZCL2013-18,Fang,Liu2017-19}.

In the past few decades, various LGPT probes have been proposed experimentally and theoretically including intermediate mass fragments (IMFs)~\cite{YGMa1995-20,MEF1969-21}, the second  moment parameter~\cite{XC1988-22,SDG1998-23}, caloric curve~\cite{TFAO2006-24,YGM2005-25,YGM2005-26,jpo1995-12}, bimodality in charge asymmetry~\cite{ODE2005-28,ALJA2008-29}, charge fluctuation of the largest fragment (NVZ)~\cite{BB2002-30,COVC1999-31}, Fisher's power-law exponent of fragments~\cite{YGM2004-32,JBLG2002-33,MHRW2010-34}, and the nuclear Zipf's law proposed by Ma~\cite{ygm1999-15,YGM1999-36}. 
Furthermore, the  derivative of total multiplicity of fragments with respect to temperature and the difference between the size of the first and second largest clusters are suggested as LGPT probes in the framework of the statistical model, canonical thermodynamical model, and statistical multifragmentation model~\cite{SGPS2017-37,SSG2018-38,PSG2018-39,CSWA2005-40,ASIN2001-41,GADA2007-42}.
Recently, studies on the sensitivity of different LGPT probes based on experimental measurements indicated that the total multiplicity derivative and NVZ might be better measures for the critical point from the final cold fragments and primary fragments~\cite{WPH2018-43}. However, discussions on the properties of these LGPT probes are still ongoing. Some of the questions raised are as follows: what is the order of the phase transition in nuclear matter? What is the finite-size effect of phase transition in nuclei?
Here, we attempt to understand the finite-size scaling effect for different nuclear LGPT probes in the framework of a quantum molecular dynamics model.

In this article, within the framework of an extended version of the isospin-dependent quantum molecular dynamics (IQMD) model, different finite-size systems are excited at temperatures ranging from 0 to 20 MeV and their de-excitation processes are obtained. Specifically, six finite-size source systems are selected, with the mass numbers $A$ = 36, 52, 80, 100, 112, and 129, and their corresponding charges $Z$ = 15, 24, 33, 45, 50, and 54, respectively, which are initialized at the same normal density based on the Fermi--Dirac distribution~\cite{Fang}. We investigate several probes, including the total multiplicity derivative ($dM_{tot}/dT$), second moment parameter ($M_{2}$), Ma's nuclear Zipf's law exponent ($\xi$), multiplicity of IMFs ($N_{IMF}$), and Fisher's power-law exponent ($\tau$), to analyze their finite-size effects, and then we apply the finite-size scaling law for  the phase transition temperatures  of different finite-size source systems to extract the critical exponent for an infinite-size source system.

The rest of this paper is organized as follows: In Section~\ref{modelformalisim}, a brief introduction for an IQMD model is provided, specifically for an IQMD version with a single thermal equilibrated source. In Section~\ref{resultanddiscussion}, the temperature evolution of different LGPT probes is discussed and finite-size scaling is applied for these probes. Finally, a conclusion is given in Section~\ref{summary}.

\section{MODEL}
\label{modelformalisim}

The IQMD model is an extended version of the quantum molecular dynamics model, which uses a many-body theory to describe the dynamical behavior of heavy-ion collisions at intermediate energies~\cite{JA1991-44,CZL1989-45,Ma_wei,Feng,Fang,ZZF2018-47}. In the IQMD model, each nucleon is represented by a Gaussian wave packet, i.e.,
\begin{equation}\label{wavefunction}
\phi_{i}\left(\mathbf{R},t\right) = \frac{1}{\left(2\pi L\right)^{3/4}}exp\left[-\frac{\left(\mathbf{R}-\mathbf{R}_{i}\right)^2}{4L}+\frac{i}{\hbar}\mathbf{R}\cdot\mathbf{P_{i}}\right],
\end{equation}
where $\mathbf{R}_{i}$ and $\mathbf{P}_{i}$ are the time-dependent coordinate and momentum, respectively.  $L$ is the Gaussian width, which is set to 2.16 $fm^{2}$ in the present work. The total effective potential of nucleons in the nuclear mean field can be expressed as
\begin{multline}\label{meanfield}
U\left(\rho,\tau_{z}\right) = \alpha \left(\frac{\rho}{\rho_{0}}\right) + \beta\left(\frac{\rho}{\rho_{0}}\right)^{\gamma} + \frac{1}{2}\left(1-\tau_{z}\right)V_{c}\\
+ C_{sym}\frac{\left(\rho_{n} - \rho_{p}\right)}{\rho_{0}}\tau_{z}
+ U^{Yuk}+U^{MDI},
\end{multline}
where $\rho$ is the total density of the nuclear system, which is calculated using the interaction density in IQMD~\cite{JA1991-44,JA1987-44,ZCL2013-18,ZGQ2011-17} as follows:
\begin{align}
\label{density}
\rho&= \sum_{i}^{A}{\rho_{i}} = \sum_{i\ne j}^{A}\frac{1}{(4\pi L)^{3/2}}exp(-\frac{(r_{i}-r{j})^{2}}{4L}),
\end{align}
and $\rho_{0}$ is the saturation density at the ground state, i.e.,  0.16 $fm^{-3}$. $\rho_{n}$ and $\rho_{p}$ are neutron and proton densities, respectively. $\alpha$, $\beta$, and $\gamma$ are parameters related to the EOS of bulk nuclear matter~\cite{JA1991-44}. $C_{sym}$ is the symmetry energy coefficient, $V_{c}$ is the Coulomb potential, and $\tau_{z}$ is the z-th component of the isospin degree of freedom for the nucleon, which equals 1 or -1 for a neutron or proton, respectively.  $U^{Yuk}$ represents the Yukawa potential, and $U^{MDI}$ represents the momentum-dependent interaction (MDI) potential, which can be written as follows:
\begin{equation}
\label{momentum_potential}
U^{MDI} = \delta\cdot ln^{2}(\epsilon\cdot(\Delta p)^{2}+1)\cdot\frac{\rho}{\rho_{0}},
\end{equation}
where $\Delta p$, which is the relative momentum, $\delta$ = 1.57 MeV, and $\epsilon$ = 500 (GeV/c)$^{-2}$ are the parameters of $U^{MDI}$ and they can be found in Refs.~\cite{JA1991-44,JA1987-44}.
Instead of the traditional collision version of the IQMD model, we use a single-source IQMD model. In this framework, the time evolution of phase space information of nucleons and fragments starts from a single thermal equilibrated source having an initial temperature and density.  This version of the IQMD model has already been described by Fang, Ma, and Zhou in Ref.~\cite{Fang}. For simplicity, the above thermal IQMD version is called ThIQMD hereafter.

In this model, the momentum distribution of the nucleons at the temperature $T$ = 0 (i.e., the nucleus is at the ground state) is calculated using the local Fermi gas approximation as
\begin{equation}
\label{fermi momentum}
P_{F}^{i}\left(\mathbf{R}\right) = \hbar[3\pi^{2}\rho_{i}(\mathbf{R})]^{\frac{1}{3}},
\end{equation}
where $\rho_{i}$ is the local density of neutrons (i = n) or protons (i = p).
For $T>0$, the momentum distribution of nucleons is replaced by the Fermi--Dirac distribution, i.e.,
\begin{equation}
\label{Fermi-Dirac}
n(e_{k}) = \frac{g(e_{k})}{e^{\frac{e_{k}-\mu}{T}}+1}
 = \frac{\frac{V}{2\pi^2}\left(\frac{2m}{\hbar^2}\right)^{3/2}e_{k}}{e^{\frac{e_{k}-\mu}{T}}+1},
\end{equation}
where $e_{k} = \frac{p^2}{2m}$ is the kinetic energy with the momentum $p$, and $V$ is the source volume. $\mu$ is the chemical potential determined by the following equation:
\begin{equation}
\label{chemical potential}
\frac{1}{2\pi^{2}}\left(\frac{2m}{\hbar^2}\right)^{3/2}\int_{0}^{\infty}\frac{e_{k}}{e^{\frac{e_{k}-\mu}{T}}+1}de_{k} = \rho.
\end{equation}

In the present work, we adopt $\alpha$ = -390.1 MeV, $\beta$ = 320.3 MeV, and $\gamma$ = 1.14, which correspond to the soft MDI EOS with the incompressibility of $K$ = 200 MeV. The fragments recognized by a coalescence mechanism in the present model are constructed with a relative distance smaller than 3.5 fm and a relative momentum lower than 300 MeV$/c$ between two nucleons.

\section{CALCULATIONS AND DISCUSSION}
\label{resultanddiscussion}

In the current work, six source systems with different masses and charge numbers but with similar $N/Z$ of approximately $\sim$ 1.3, i.e., (A,Z) = (36, 15), (52, 24),  (80, 33),  (100, 45),  (112, 50), and (129, 54), are excited at an initial temperature varying from 0 to 20 MeV. The de-excitation process can be studied using the ThIQMD model ~\cite{Fang}.

\subsection{Phase transition temperature for finite-size nuclear sources}

\begin{figure}
\includegraphics[width=9.5cm]{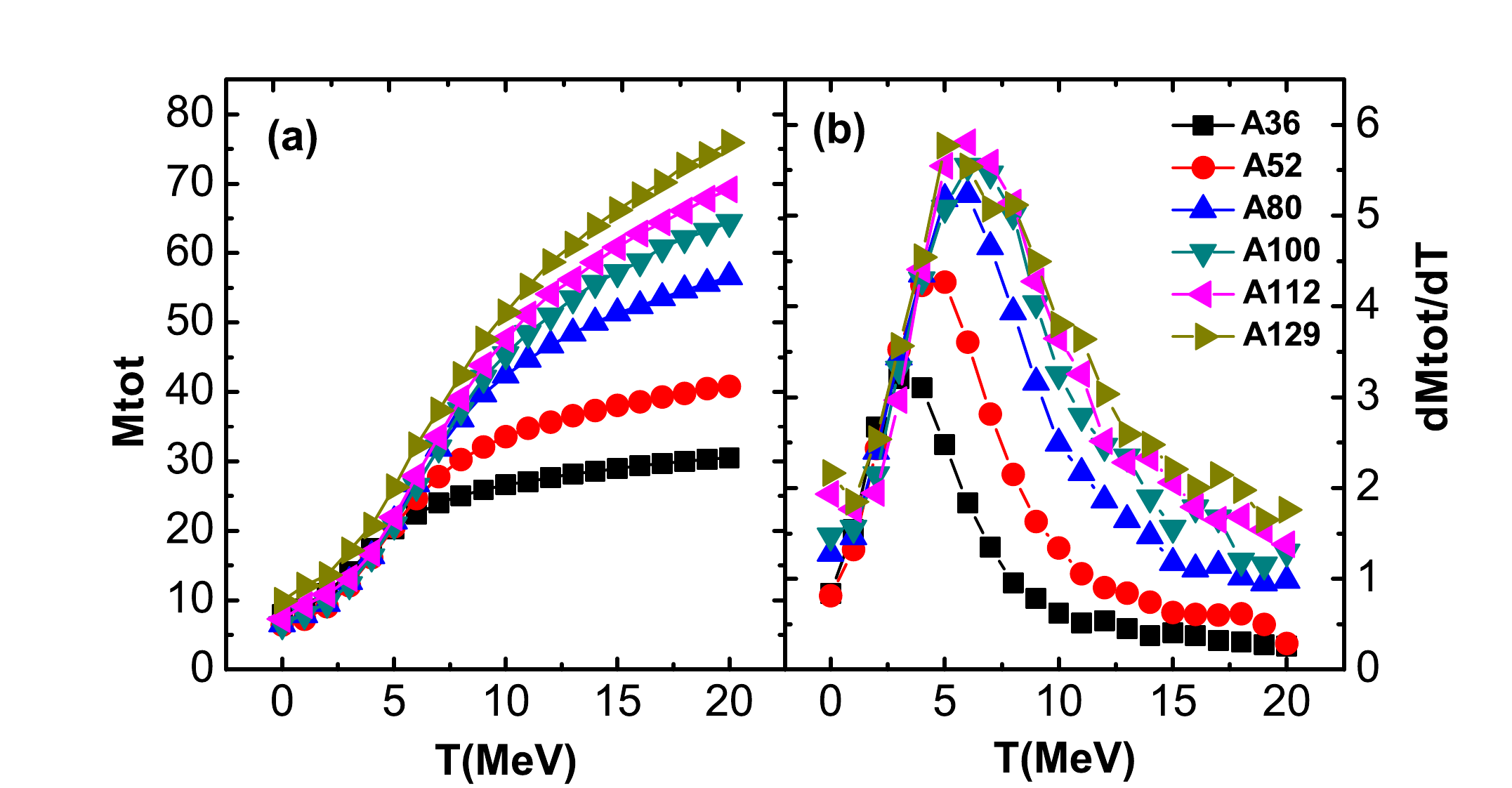}
\caption{(Color online) Temperature dependences of the total multiplicity (a) and its derivatives versus source temperature (b) in the ThIQMD model for  different source-size systems.}
\label{Mtot}
\end{figure}

After the model completes the dynamic evolution process at the initial temperature, the final state observables can be obtained when the system is at freeze-out state.  Fig.~\ref{Mtot}(a) shows the relationship between the total particle multiplicity ($M_{tot}$) and the source temperature, which is obtained in the final state. The relationship between the derivative of $M_{tot}$ w.r.t. source temperature and the source temperature is also obtained, which is shown in Fig.~\ref{Mtot}(b). It can be observed that the total multiplicity increases with the source temperature. Moreover, the derivative of total multiplicity w.r.t. temperature increases initially and thereafter decreases after peaking at a certain temperature. This is similar to the results provided by Mallik {\it et al}~\cite{SGPS2017-37}. As suggested in Refs.~\cite{SGPS2017-37,SSG2018-38}, the temperature corresponding to this peak is the apparent phase transition temperature, which is denoted by $T_A$. From Fig.~\ref{Mtot}, the phase transition temperature $T_A$ increases with the size of the source and tends to saturate for larger systems.

\begin{figure}
\includegraphics[width=9.5cm]{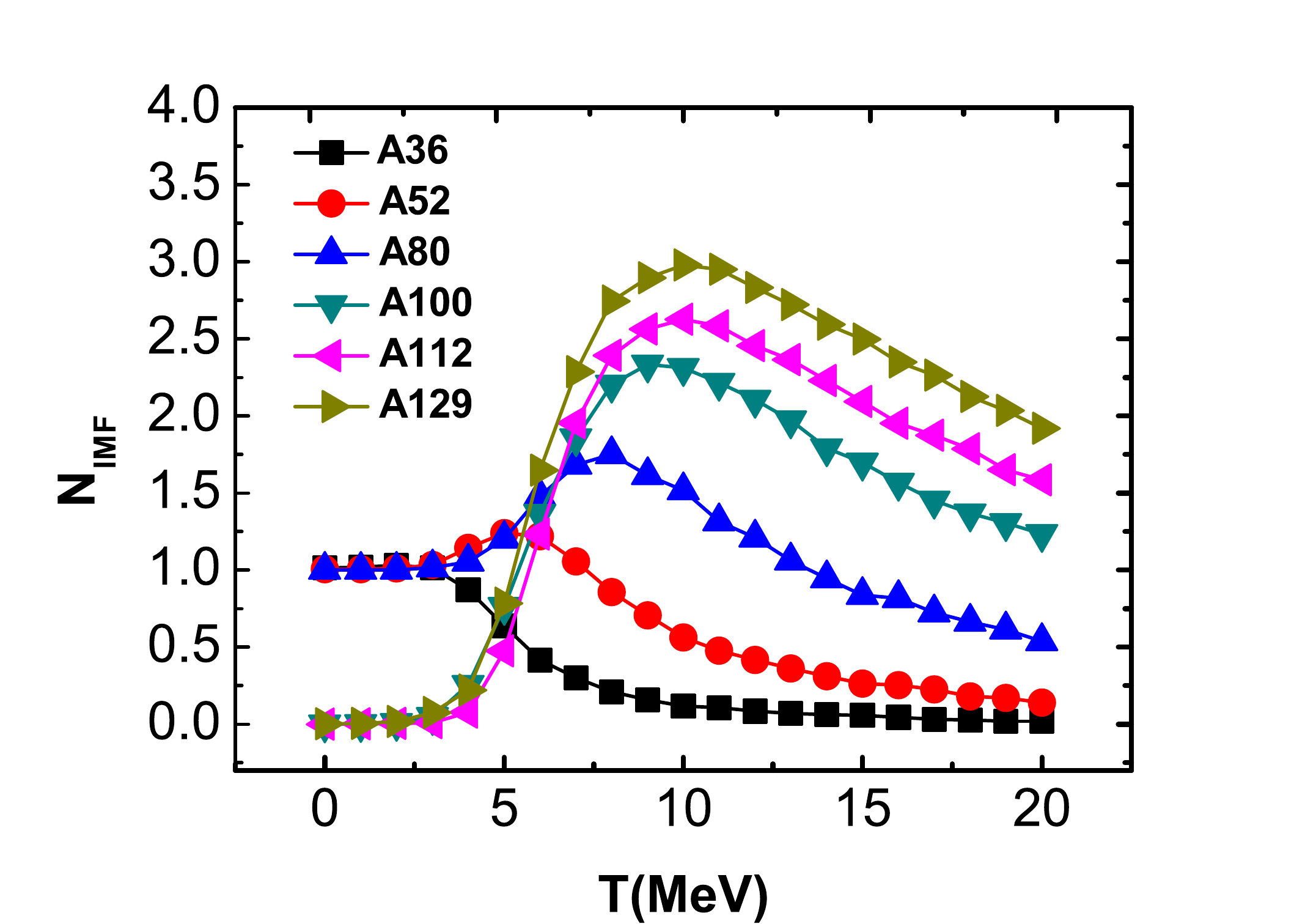}
\caption{(Color online) Temperature dependence of the IMF multiplicity  in the ThIQMD model for different source-size systems.}
\label{Mimf}
\end{figure}

Fig.~\ref{Mimf} shows the evolution of the number of IMFs ($N_{IMF}$) with the source temperature, where the IMF is defined as a fragment whose charge number ($Z$) is larger than 3 and smaller than the charge number of the source. It is observed that, as the temperature increases, the number of IMFs has an increasing trend and then a decreasing trend. There exists a broad peak in the temperature evolution, which is consistent with the previous experimental results of Y. G. Ma {\it et al.}~\cite{YGMa1995-20}. The temperature corresponding to the peak position represents the onset point of LGPT. At the phase transition temperature, we also observe that $N_{IMF}$ increases with an increase in the source size.

\begin{figure}
\includegraphics[width=9.5cm]{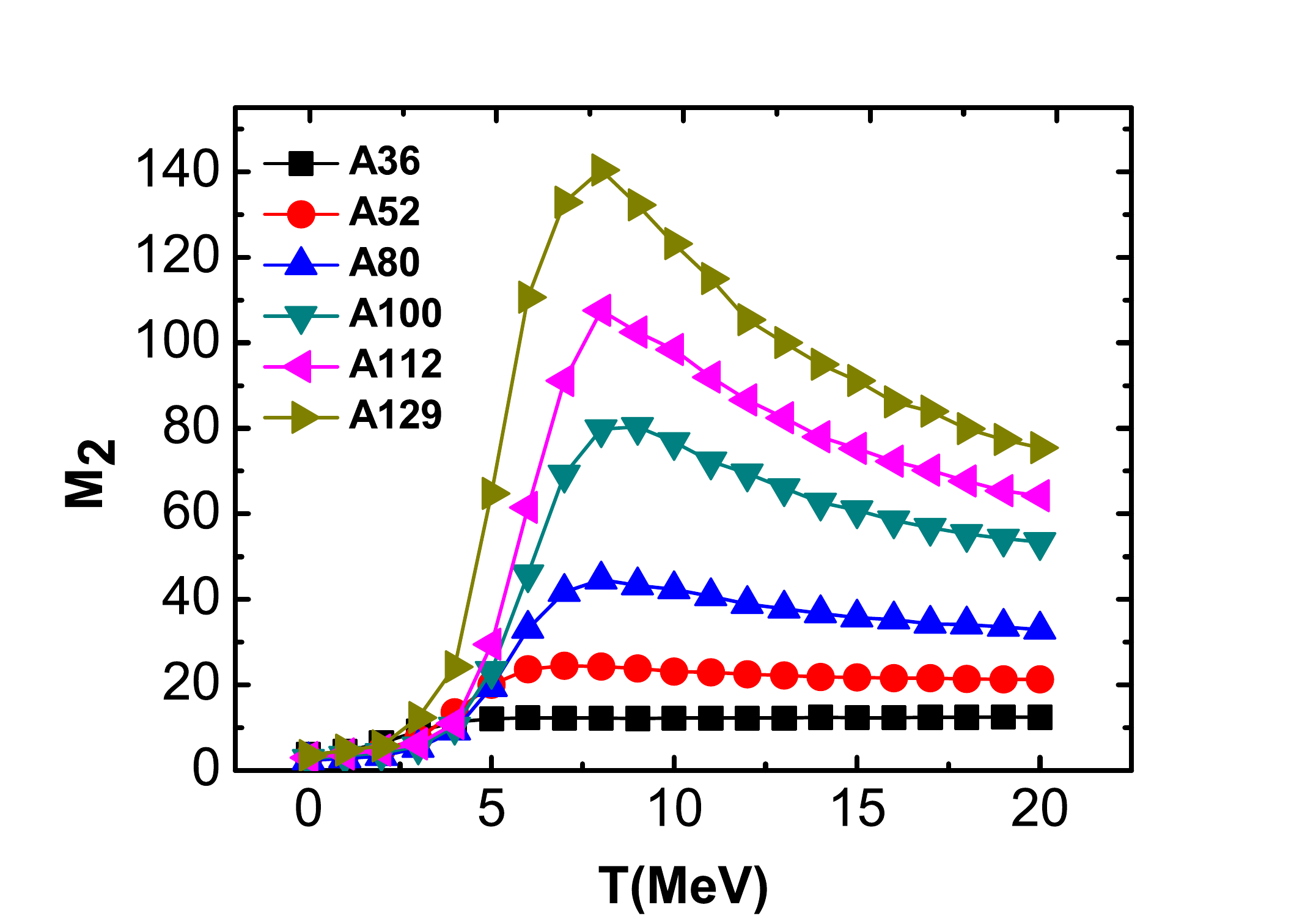}
\caption{(Color online)  Same as Fig.~\ref{Mimf} but for  $M_{2}$.}
\label{M2}
\end{figure}

Through application of the information of fragments, the value of the second moment $M_2$ can be obtained, which could be employed to search for the phase transition region~\cite{XC1988-22,SDG1998-23}. The general definition of the $k$-th moment is
\begin{equation}\label{k-th moment}
M_{k} = \sum_{Z_{i}\neq Z_{max}}Z_{i}^{k}\cdot n_{i}\left(z_{i}\right).
\end{equation}
The results are shown in Fig.~\ref{M2}. The figure shows that $M_2$ increases first and then decreases with the temperature, which is similar to that observed for the derivative of total multiplicity w.r.t. temperature and the multiplicity of IMFs.

\begin{figure}
\includegraphics[width=9.5cm]{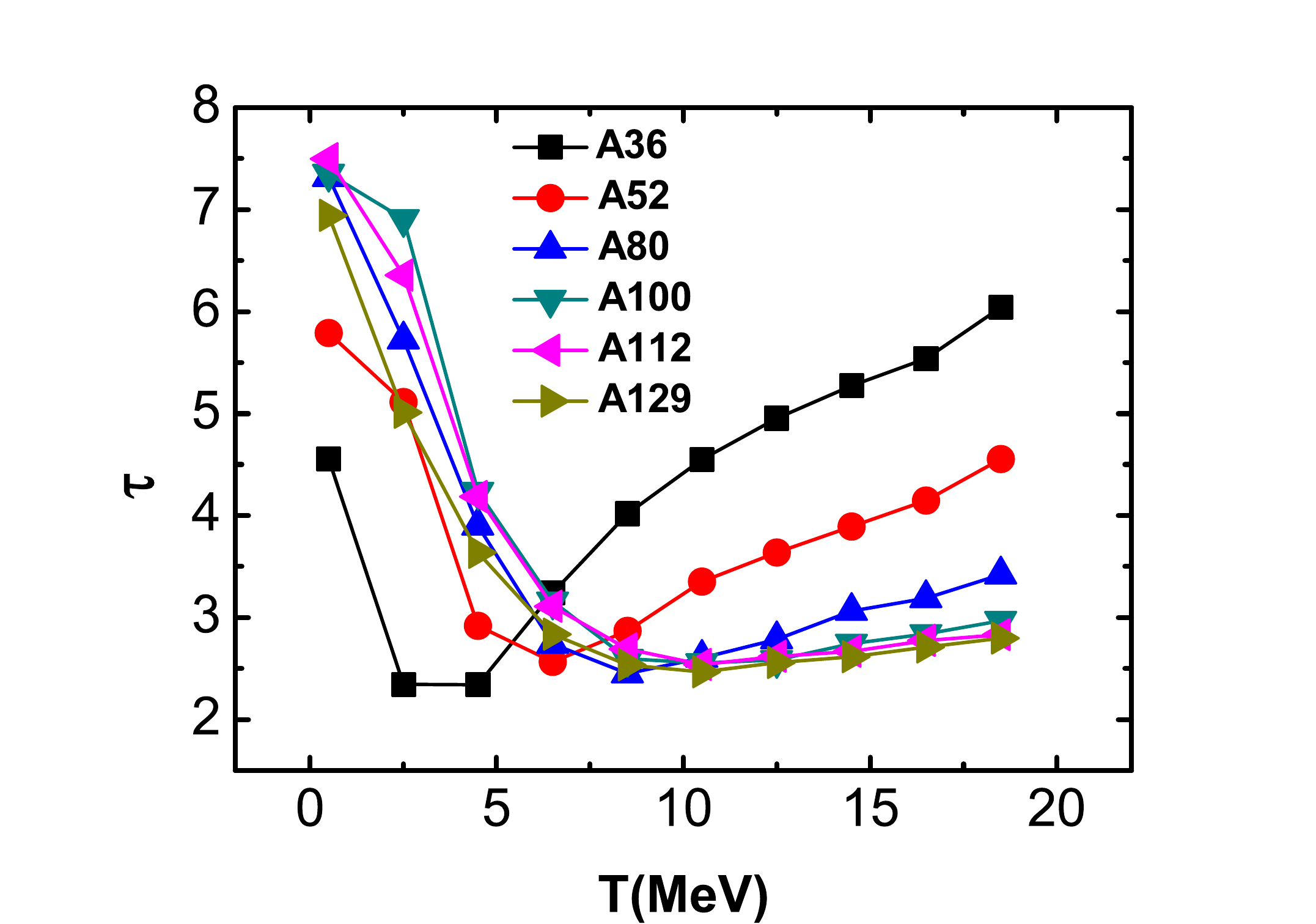}
\caption{(Color online) Same as Fig.~\ref{Mimf} but for  Fisher's power-law exponent ($\tau$).}
\label{Fisher}
\end{figure}

Fisher's droplet model is extensively applied to the study of multifragmentation, and its power-law exponent ($\tau$) is applied for locating the phase transition point where its value is usually approximately $2\sim3$~\cite{MHRW2010-34}. In Fisher's droplet model ~\cite{MEF1969-21}, the fragment mass yield distribution $Y(A)$ can be expressed as
\begin{equation}\label{taulaw}
Y(A) = Y_{0}A^{-\tau}
\end{equation}
in a suitable mass or charge region.
As the power-law exponent of mass or charge distributions behaves in a similar manner, as shown in Ref.~\cite{ROAS2002-48}, we employed the power-law fit for charge distribution in the range $Z$ = 2 $\sim$ 7 for different source-size systems, which eliminates the fluctuation impact of large or small fragments. The temperature dependence of the Fisher's exponent ($\tau$), which is obtained from the $Z$ distribution, is shown in Fig.~\ref{Fisher}. It is evident that $\tau$ has a minimum value between 2 and 3, consistent with the results obtained by Y. G. Ma {\it et al.}~\cite{YGMa1995-20} and many others. Interestingly, the Fisher's exponent ($\tau$) increases faster for lighter sources than for heavier sources after the minimum.

\begin{figure}
\includegraphics[width=9.5cm]{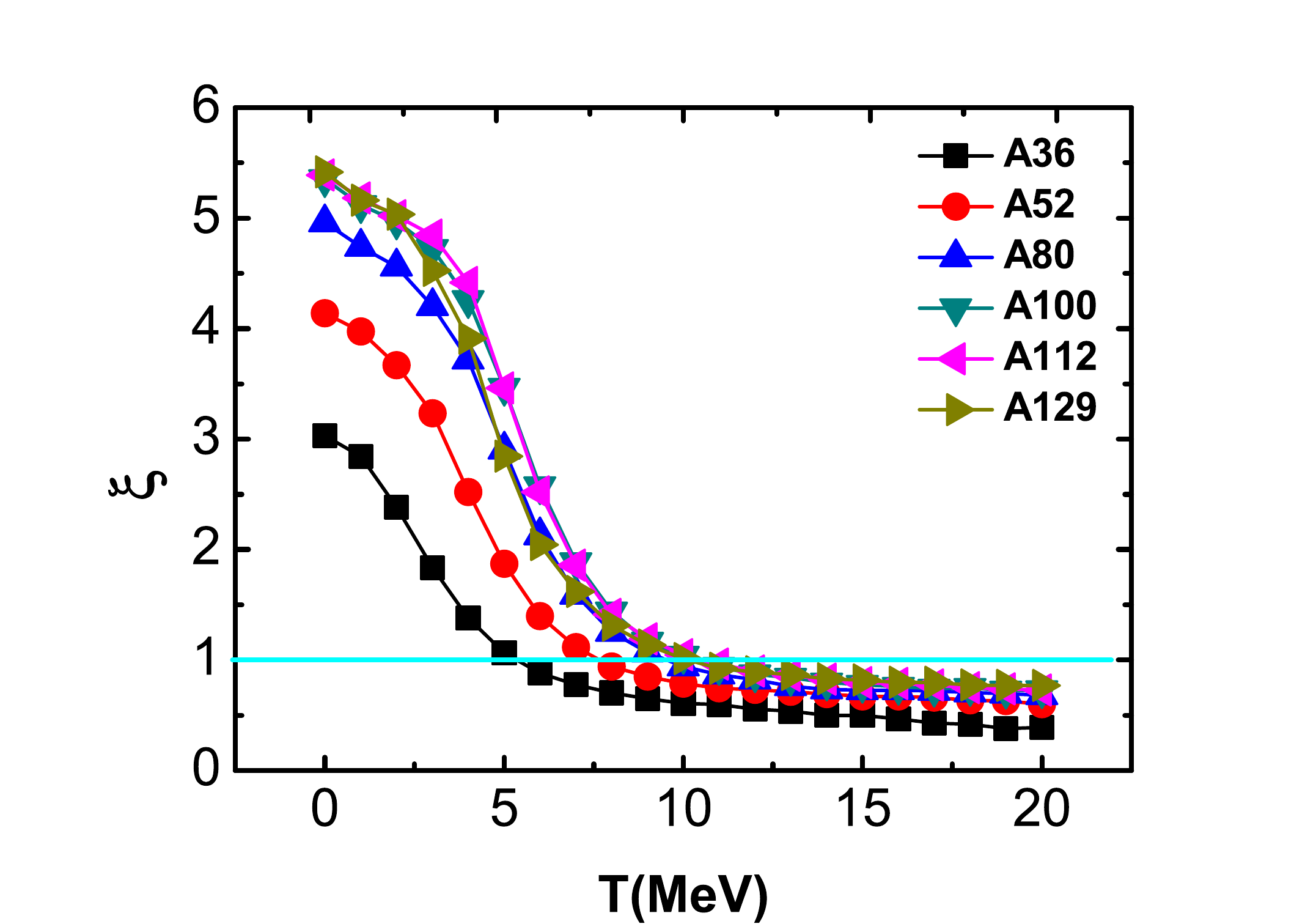}
\caption{(Color online) Same as Fig.~\ref{Mimf}  but for Ma's nuclear Zipf's law exponents  ($\xi$). The cyan horizontal line represents $\xi$ equal to 1.}
\label{Zipf}
\end{figure}

The nuclear Zipf's law was first proposed by Y. G. Ma by defining a nuclear Zipf-type plot~\cite{ygm1999-15,YGM1999-36}, where the average cluster size is employed as a function of rank and the resultant distribution is fitted with a power law given by
\begin{equation}\label{Zipflaw}
\left<Z_{rank}\right>\propto rank^{-\xi},
\end{equation}
where $\xi$ is Ma's Zipf's law exponent and $rank $= $i$ for the $i-$th largest fragment. In nuclear fragmentation, Ma's Zipf's law reflects the hierarchy distribution of the fragment, which provides a valuable method to search for the LGPT in finite systems~\cite{ygm1999-15,YGM1999-36}. When $\xi$$\sim$1, Ma's nuclear Zipf's law is satisfied.  The values of Ma's Zipf's law exponent $\xi$, which describe the fragment rank distribution of different source-size systems, are plotted in Fig.~\ref{Zipf}. As the value of $\xi$ equals 1, the Ma's nuclear Zipf's law is satisfied and the corresponding phase transition temperature of different source-size systems can be extracted.

\begin{figure}
\includegraphics[width=9.5cm]{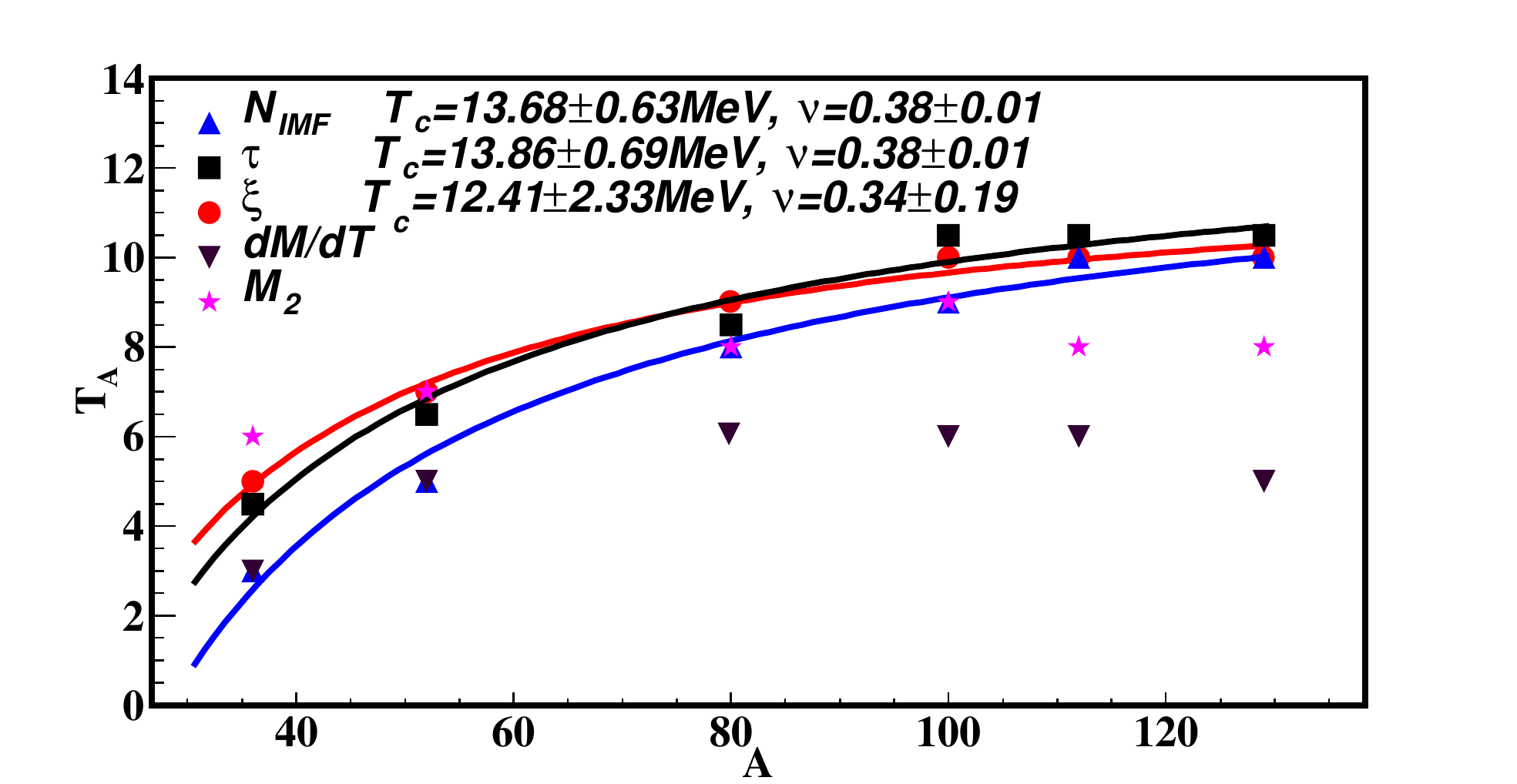}
\caption{(Color online) Mass dependence of the phase transition temperatures ($T_A$), which are obtained with different probes.
For temperatures deduced from $\xi$ (red line), $\tau$ (black line), and $N_{IMF}$(blue line), the critical temperature of an infinite-size source system and the critical exponent $\nu$ are obtained by fitting using Eq.~(\ref{T-scaling}).
 It is observed that, for an infinite-size source system, the critical temperature is 13.32 $\pm$ 1.22 MeV on average, and $\nu$ is 0.37 $\pm$ 0.07.}
\label{TA}
\end{figure}

To evaluate the sensitivity of the above mentioned probes to the size of the system, the phase transition temperature values obtained with different probes are shown as a function of the size of the system. From Fig.~\ref{TA}, it can be observed that, for the IMF multiplicity ($N_{IMF}$), Fisher's exponent ($\tau$), and Ma's Zipf's law exponent  ($\xi$), the obtained phase transition temperature values change considerably as the source size increases. However, the values of the phase transition temperature derived from $M_{2}$ and the derivative of the total multiplicity w.r.t. temperature $dM_{tot}/dT$ vary slowly with the source size. In this context, we assume that the finite-size effect should be considered when probing the critical temperature ($T_c$), which is the phase transition temperature for infinite nuclear matter, employed by the approaches of the IMF multiplicity, Fisher's exponent, and Ma's Zipf's law exponent.

\subsection{Critical temperature for infinite-size nuclear source}

In the above examination, the source sizes of systems are finite and the deduced phase transition temperatures are valid only for specific finite-size nuclei. However, from Fig.~\ref{TA}, we observe that the phase transition temperatures obtained with the probes, namely $N_{IMF}$, $\tau$, and $\xi$, have a strong correlation with the source size. It is well known that finite-size scaling is an important factor in studying the phase transition of infinite matter. In this context, we shall use the finite-size scaling law to deduce the critical temperature as the source size tends to infinity \cite{YGM1999-36,Anto,Pal,ChenXS}.
We apply the extrapolation rule
\begin{equation}
\left|T_{c} - T_{A} \right|\propto A^{-\frac{1}{3\nu}}
\label{T-scaling}
\end{equation}
to estimate the critical value $T_{c}$ when the nuclear LGPT takes place as $A$ $\rightarrow$ $\infty$ and extract the $\nu$ parameter, which is a critical exponent for correlation length. Usually, for a
three-dimensional (3D) percolation class,  $\nu$ is  0.9;  for the 3D
Ising class or liquid--gas class,  $\nu$ is equal to 0.6 \cite{Muller}.
The fitting results are shown in Fig.~\ref{TA},
indicating that the above finite-size scaling law
is valid  for the phase transition temperatures deduced from the IMF multiplicity, charge distribution, and Ma's nuclear Zipf's law. The fitting results demonstrate that, for different phase transition probes, the best-fitted critical temperatures $T_c$ for an infinite nuclear system are consistent with each other, i.e., approximately  13.32 $\pm$ 1.22 MeV on average, which are close to the experimentally determined value ($T_c$ = 16.6 $\pm$ 0.86 MeV) in symmetric nuclear matter, obtained by Natowitz {\it et al.} \cite{ygm2002-16} as well as a recent first-principle prediction for the LGPT critical temperature ($T_c$ = 16 $\pm$ 2 MeV) in symmetric nuclear matter employing
both two- and three-nucleon chiral interactions ~\cite{Rios}. Furthermore,
the critical exponents for the correlation length ($\nu$) for the above three probes have similar values, 0.37 $\pm$ 0.07, indicating phase transition behavior
~\cite{PHW1978-49,ELP2011-50}. The present $\nu$
value appears to approach the value of 3D Ising
or liquid--gas class rather than the 3D percolation class, which indicates that the nuclear disassembly in this model belongs to the 3D liquid--gas universality class.
 However, $T_A$ values deduced from the total multiplicity derivative ($dM_{tot}/dT$) and the second moment parameter ($M_{2}$) do not follow the finite-size scaling law well. Thus, they might not be good probes for the critical behavior of nuclear LGPT.

 To further verify the above finite-size scaling law, we plot ln$|T_c - T_A |$ as a function of -$\frac{1}{3\nu}$lnA where $T_c$ = 13.32 MeV and $\nu$ = 0.37 are  assumed, and $T_A$ represents the phase transition temperatures obtained from $\xi$, $\tau$, and $N_{IMF}$. In Fig. 7, a line with the slope = 1 is also plotted. All three curves appear linear  with a slope of 1 in  the double-ln coordinate and are close to each other, which illustrates that the deduced $T_c$ and $\nu$ are the correct values.

\begin{figure}
\includegraphics[width=9.5cm]{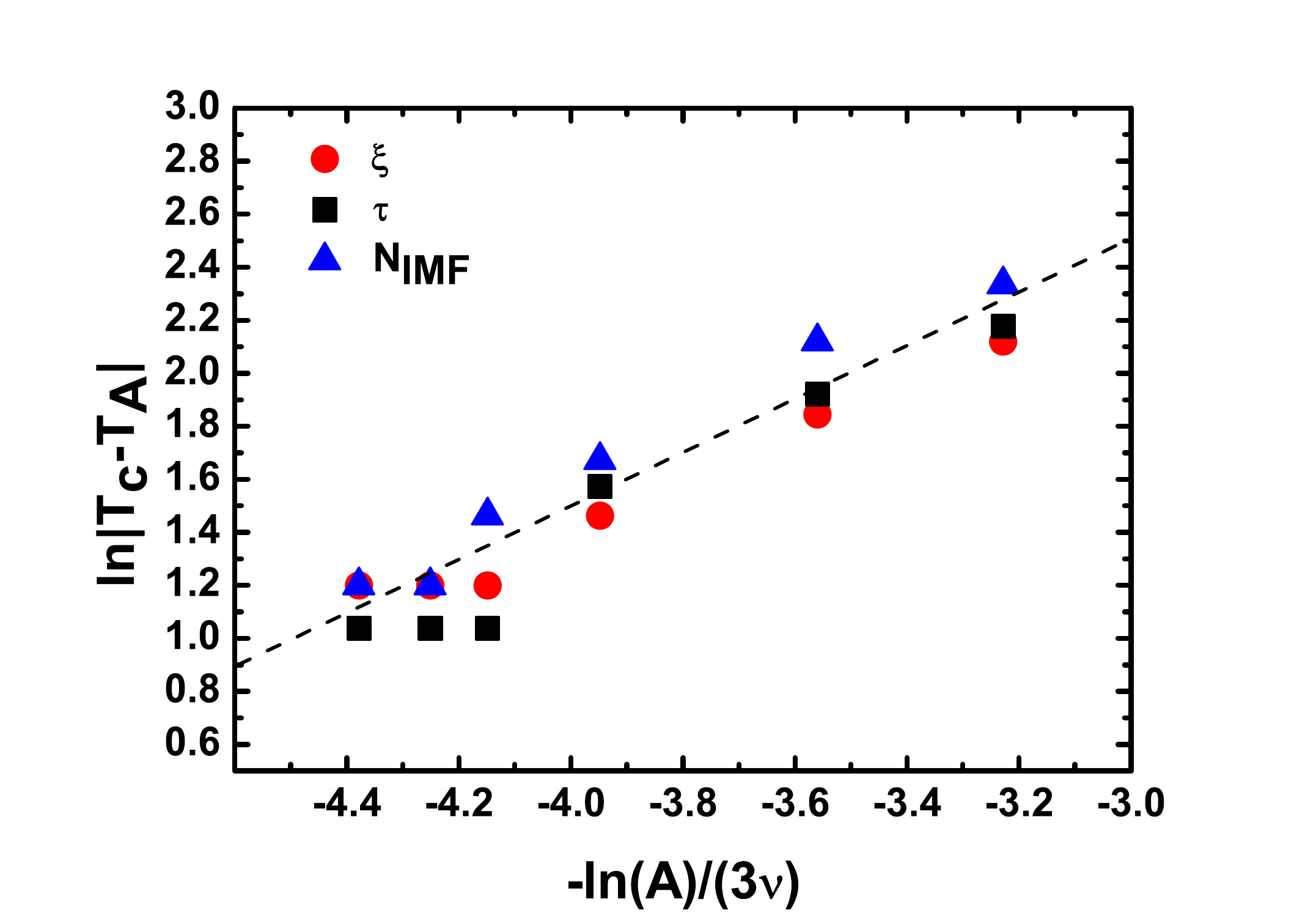}
\caption{(Color online) Plot of ln$|T_c - T_A |$ as a function of -$\frac{1}{3\nu}$lnA where $T_c$ = 13.32 MeV and $\nu$ = 0.37 are assumed. The slope of the dashed line is 1.
}
\label{miu}
\end{figure}

\section{SUMMARY}
\label{summary}

In summary, the de-excitation processes of six finite-size systems at the same initial normal density have been investigated using a single thermal source version of the IQMD model (ThIQMD). Their temperatures at the point of LGPT were obtained by checking different probes, such as the IMF multiplicity ($N_{IMF}$), Fisher's power-law exponent of charge distribution ($\tau$), Ma's Zipf's law exponent ($\xi$), derivative of total multiplicity w.r.t. temperature ($dM_{tot}/dT$), and second moment parameter ($M_{2}$). The results demonstrate that the phase transition temperatures obtained from all these probes have a positive correlation with the source system size and tend to saturate. Moreover, judging from the finite-size scaling law, the phase transition temperatures obtained from  $N_{IMF}$, $\tau$, and $\xi$ follow the same scaling law well, whereas those from $dM_{tot}/dT$ and $M_{2}$ do not. Thus, the IMF multiplicity ($N_{IMF}$), Fisher's power-law exponent, and Ma's nuclear Zipf's law ($\xi$) are good probes for estimating the critical behavior of phase transition, whereas the total multiplicity derivative and second moment are not.
Based on the finite-size scaling law, the deduced critical temperature is approximately 13.32 $\pm$ 1.22 MeV and the $\nu$ exponent is 0.37 $\pm$ 0.07 for the infinite nuclear system with $N/Z \sim 1.3$, and the value of {$\nu$} is close to the value of 3D Ising or liquid--gas universal classes.

\begin{acknowledgments}

We would like to thank Dr. Zhenfang Zhang for his useful discussions. This work is supported by the National Natural Science Foundation of China under Contract Nos. 11421505 and 11890714, the National Key R$\&$D Program of China under Contract No. 2018YFA0404404, the Key Research Program of Frontier Sciences of the CAS under Grant No. QYZDJ-SSW-SLH002, and the Strategic Priority Research Program of the CAS under Grant No XDPB09 and XDB16.

\end{acknowledgments}

\end{CJK*}
\end{document}